\documentclass[mathleft,fleqn,%
]{an}
\usepackage{graphicx}
\usepackage[varg]{txfonts}
\usepackage{natbib}
\bibpunct{(}{)}{;}{a}{}{,}
\begin{document}

% The following seven commands are intended for editorial usage and
% should be ignored by the author(s).
\Pagespan{1}{}% Document's page range. 
% If second parameter is left empty, the last page is computed
% automatically.
\Yearpublication{2017}%
\Yearsubmission{2017}%
\Month{0}%   
\Volume{999}%  
\Issue{0}% 
\DOI{asna.201400000}% 

\title{
        Rotation--differential rotation relationships for  late-type single and binary stars from Doppler imaging%\,\thanks{Data from STELLA}
}

\author{Zs. K\H{o}v\'ari\inst{1}\fnmsep\thanks{Corresponding author:
        {kovari@konkoly.hu}}
% Example for footnote, note the usage of the \texttt{fnmsep} command
% as separator between institute number and footnote mark}
\and  K. Ol\'ah\inst{1}
\and  L. Kriskovics\inst{1}
\and  K. Vida\inst{1}
\and  E. Forg\'acs-Dajka\inst{2}
\and  K.\,G. Strassmeier\inst{3}
}
\titlerunning{Rotation--differential rotation relationships for late-type stars}
\authorrunning{Zs. K\H{o}v\'ari et al.}
\institute{
Konkoly Observatory, Research Center for Astronomy and Earth Sciences, Budapest, Hungary
\and 
E\"otv\"os University, Department of Astronomy, Budapest, Hungary
%DARK, Niels Bohr Institute, University of Copenhagen, Denmark
\and 
Leibniz-Institute for Astrophysics Potsdam (AIP), Germany}

\received{XXXX}
\accepted{XXXX}
\publonline{XXXX}


\abstract{From our sample of spotted late-type stars showing surface differential rotation we find that the relationship between the rotation period and the surface shear coefficient $\alpha=\Delta\Omega/\Omega_{\rm eq}$ is significantly different for single stars compared to members in close binaries. Single stars follow a general trend that $\alpha$ increases with the rotation period. However, differential rotation of stars in close binary systems shows much weaker dependence on the rotation, if any, suggesting that in such systems tidal forces operate as a controlling mechanism of differential rotation.}

\maketitle

\section{Introduction}

Stellar dynamos working in late-type stars generate strong magnetic fields, which, at the end, are manifested in activity phenomena such like starspots. Differential rotation of spotted stars with convective envelopes is of utmost importance in understanding how the dynamo mechanism (re)generates large scale toroidal fields by the so-called $\Omega$-effect. However, magnetic dynamos work diversely in different types of stars.

We have learned that rotation is the main driving force that can sustain the dynamo, still, it is not known what kind of relationship exists between rotation and differential rotation, if any.
%{\bf 
In principle, with increasing rotation rate the role of differential rotation is getting less significant, i.e., according to the mean field theory, a transition may be hypothesized from $\alpha\Omega$ 
and $\alpha^2\Omega$-type dynamos to $\alpha^2$ dynamos \citep{2015SSRv..196..303B}. 
% }
While in rapidly rotating young late-type (G--K) stars supposedly $\alpha^2\Omega$-type dynamos work, in fully convective low-mass  stars or brown dwarfs a pure $\alpha^2$-type dynamo may operate, wherein the helical turbulence ($\alpha$-effect) is dominant and the $\Omega$-effect is basically negligible \citep{2006A&A...446.1027C}. For the effect of the Rossby number on the dynamo in fully convective M-dwarfs and the presumed transition from $\alpha^2$ dynamos to $\alpha\Omega$-type see \citet{2016ApJ...833L..28Y}.

In RS\,CVn-type binary systems tidal coupling is responsible for maintaining fast rotation. Also, the gravitational influence of a close companion yields different physical conditions  inside a differentially rotating convective bulk, which may imply that dynamos work differently in single stars compared to components of RS\,CVn systems \citep[][]{2012IAUS..282..197K}. Although, the related background theory is still too complex in its predictive power, for now the observational database is wide enough to study the relationship between the rotation rate and the differential rotation.

Starspots are proved to be useful tracers for measuring surface differential rotation on either single stars or members in binary systems (see, e.g., \citealt{1997MNRAS.291....1D,2007AN....328.1075W,2012A&A...539A..50K}; and see also the review by \citealt{2009A&ARv..17..251S}).
In this paper we aim to analyze how the differential surface shear depends on the rotation, also, how the relationship is affected by the binarity. We collect surface shear coefficients from the literature, based on the most reliable Doppler imaging and Zeeman--Doppler imaging studies from the past two decades. Observations have already proved the existence of antisolar-type differential rotation \citep[see, e.g.,][and their references]{2014SSRv..186..457K}, i.e., when the rotation is the slowest at the equator and increases with latitude towards the pole. In Sect.~\ref{disc} we suggest a possible scenario which may yield such a peculiar surface rotation pattern on single K-giants arriving at the red giant branch (RGB).

\section{Measuring surface differential rotation}

Stellar surface differential rotation laws are generally written in a quadratic form of
\begin{equation}
  \label{drlaw}
\Omega(\beta)=\Omega_{\rm eq}(1-\alpha\sin^2\beta),
\end{equation}
where $\Omega(\beta)$ is the angular velocity at $\beta$ latitude, $\Omega_{\rm eq}$
is the angular velocity at the equator, while the dimensionless relative surface shear $\alpha$ is expressed as $\alpha=\Delta\Omega/\Omega_{\rm eq}$, where
$\Delta\Omega=\Omega_{\rm eq}-\Omega_{\rm pol}$ is the absolute shear. In this context $\alpha>0$ stands for solar-type differential rotation, when the angular velocity is maximum at the equator, while $\alpha<0$ means antisolar differential rotation, when the equatorial belt rotates the most slowly.

When measuring surface differential rotation of late-type stars, basically two methods are considered, namely the cross-correlation of subsequent
Doppler (or Zeeman--Doppler) images and the sheared image method, also known as parametric imaging. 
For the cross-correlation technique 
two subsequent image reconstructions are needed and the rotationward cross-correlation function map obtained from the images may reveal the differential rotation pattern.
However, this pattern can easily be blurred by rapid surface evolution (e.g., emerging a new spot, merging and dissolving spots).
On the other hand, when having more than two Doppler images in time series, such unwanted effects can be reduced by averaging all the available cross-correlation maps, yielding a more reliable result. For the detailed description of this average cross-correlation method see, e.g. \citet[]{2015A&A...573A..98K} and the references therein. 
The sheared image method \citep[e.g.][]{2000MNRAS.316..699D,2002MNRAS.334..374P}, however, works even when having only one single Doppler image. The surface shear is incorporated in the reconstruction process as a predefined parameter while image reconstructions are carried out for a meaningful range of the surface shear--equatorial rotation parameter plane. Each Doppler reconstruction has a goodness-of-fit value which may help to find the most probable surface rotation law. We note that this parametric imaging works for less data compared to the cross-correlation technique, but, maybe at the cost of reliability; for critical remarks on this issue see, e.g., \citet{2014SSRv..186..457K} and their references. In addition, \citet{2014IAUS..302..198K} demonstrated that a large polar cap, frequently detected in rapid rotators, could also yield false measure of the surface shear when applying parametric imaging.

\section{The collected observational sample}\label{sample}

In this paper we focus on spotted late-type stars, i.e., solar-type or later classifications (G--K--M), confining to the evolutionary phase from zero age main sequence up to the RGB. In this evolutionary phase, dynamos
in single stars are thought to be influenced mostly by the rotation and spectral type through the related convection zone depth.
The dynamo process, however, is expected to be perturbed by tidal effects when having a close companion star. Therefore, when setting up our target list we focus on the single--binary distinction as well.

Our observational sample is based on Doppler imaging and Zeeman--Doppler imaging studies from the literature. Differential rotation measurements from Stokes V, i.e., from tracing magnetic features, usually yield higher shear values than the Stokes I results of the same target (cf., e.g., \citealt[][their Table~7]{2004MNRAS.348.1175P}; \citealt[][their Fig.~11]{2011MNRAS.413.1949W}). This is partially explained by the different anchoring depths of surface spots and magnetic tracers, however, the given errors of the corresponding shear values are higher as well, which suggests more uncertainty for the Stokes V results compared to Stokes I. Therefore, when both available, we preferred choosing Stokes I measurements, which are more compatible with the results from traditional Doppler imaging. In this study, however, we neglect rapidly rotating pre-main sequence stars, for which differential rotation values often show unexpected irregularities \citep[e.g.][]{2011MNRAS.413.1939M,2011MNRAS.413.1949W} possibly due to the vivid evolutionary phase or other unidentified effects. Although, seasonal changes of the rotation period derived from long-term photometric datasets or  the spread in the rotation frequency
could also be considered as clues for the differential rotation \citep[e.g.][]{2014MNRAS.441.2744V,2016MNRAS.461..497B}, however, photometric analysis enables only a rough estimation of $\alpha$ (without sign), therefore we do not include those results in our list.

Table~\ref{Tab1} lists the selected targets
together with their basic properties such as $P_{\mathrm{rot}}$,  $T_{\mathrm{eff}}$, luminosity class and binarity along with the measured absolute and relative surface shear values, i.e., $\Delta\Omega$ and $\alpha=\Delta\Omega/\Omega_{\mathrm{eq}}$, respectively, and the corresponding references. 
The list consists of 37 differential rotation measurements from 24 stars, including 8 single dwarfs, 6 single or effectively single giants, and 5 subgiant and 5 giant members in close binary systems. In the sample 26 detections reflect solar-type differential rotation, while 11 detections are of antisolar type. This latter was found mainly for giants with relatively long rotational periods, being either single stars (e.g., DI\, Psc, V1192\,Ori) or members in RS CVn systems (e.g., $\sigma$\,Gem, HK\,Lac).
For some targets the differential rotation was determined independently either by different authors, different methods, and/or for different epochs. Usually, these
multiple determinations are in good agreement, but there are also divergent or even contradictory results. For instance, a quite peculiar result was reported for IM\,Peg by \citet{2007AN....328.1047M},
where the authors used a 2.7 year-long observational dataset to extract 22 differential rotation measurements. It was found, that the $\Delta\Omega$ surface shear had fluctuated between about $-0.7^{\circ}$/d and +2.1$^{\circ}$/d, i.e., between antisolar and solar-type, which is quite dubious. Unsurprisingly, among other possible explanations, the authors considered that the error bars of their method was underestimated, i.e., the scatter of their detections reflected rather an ultimate error instead of real fluctuations. Therefore, in this special case we use their grand average shear value. In this context we note, that in most cases the error bars of the $\alpha$ values given in Table~\ref{Tab1} are significantly underestimated and the more realistic
relative errors should be around 30--40\% \citep{2004PADEU..14..221K,2015A&A...573A..98K}.

\begin{table*}[t!]
\centering
\caption{Surface differential rotation parameters from Doppler imaging studies.}
\label{Tab1}
 \begin{tabular}{lcccccccl}
\hline\noalign{\smallskip}
Star & $P_{\mathrm{rot}}$ [d] & $\Omega_{\mathrm{eq}}$ [$^{\circ}$/d] & $\Delta\Omega$  [$^{\circ}$/d] & $\alpha=\Delta\Omega/\Omega_{\mathrm{eq}}$ & $T_{\mathrm{eff}}$ [K] & type$^a$ & method$^b$ & reference$^c$ \\
\hline\noalign{\smallskip}
BO\,Mic & 0.38 & $947.44 \pm 0.034$ & $1.891 \pm 0.172$ & $0.0020 \pm 0.0002$ & 4890 & V,s & sim & B05 \\
LO\,Peg & 0.42 & $851.415 \pm 0.155$ & $2.005 \pm 0.401$ & $0.0024 \pm 0.0005$ & 4600 & V,s & sim  & BCL05 \\
HK\,Aqr & 0.43 & $834.97 \pm 0.132$ & $0.2808 \pm 0.115$ & $0.0003 \pm 0.0001$ & 3700 & V,s & sim & BJC04\\
AB\,Dor & 0.51 & $701.459 \pm 0.014$ & $3.2315 \pm 0.701$ & $0.0046 \pm 0.001$ & 5000 & V,s & ccf & DC97 \\
V557\,Car & 0.57 & $638.218 \pm 0.458$ & $1.43 \pm 0.860$ & $0.0022 \pm 0.001$ & 5800 & V,s &  sim & MWC05 \\
V557\,Car & 0.57 & $641.713 \pm 0.573$ & $8.02 \pm 0.573$ & $0.0123 \pm 0.0008$ & 5800 & V,s &  sim & MWC05 \\
LQ\,Hya & 1.60 & $224.9 \pm 0.02$ & $0.573 \pm 0.172$ & $0.0025 \pm 0.001$ & 5019 & V,s & sim & DCP03\\
LQ\,Hya & 1.60 & $224.9 \pm 0.02$ & $11.12 \pm 1.15$ & $0.0494 \pm 0.005$ & 5019 & V,s & sim & DCP03 \\
LQ\,Hya & 1.60 & $225.287 \pm 4.01$ & $1.261 \pm 0.458$ & $0.0056 \pm 0.0022$ & 5070 & V,s & ccf & KSG04 \\
EI\,Eri & 1.95 & $188.0 \pm 0.2$ & $7.0 \pm 1.9$ & $0.036 \pm 0.01$ & 5500 & IV,b & ccf &  KWF09 \\
FK\,Com & 2.40 & $149.769 $ & $0.015\pm 0.030$ & $0.0001 \pm 0.0002$ & 5000 & III,s & ccf & KBH00 \\
V711\,Tau & 2.84 & $127.305\pm 0.275 $ & $-0.083 \pm 0.467$ & $-0.001 \pm 0.004$ & 4750 & IV,b & sim & DCP03 \\
V711\,Tau & 2.84 & $127.323\pm 0.241 $ & $1.215 \pm 0.464$ & $0.010 \pm 0.004$ & 4750 & IV,b & sim & DCP03 \\
V711\,Tau & 2.84 & $127.432\pm 0.023 $ & $0.871 \pm 0.047$ & $0.007 \pm 0.001$ & 4750 & IV,b & sim & PDV04 \\
V1794\,Cyg & 3.30 & $110.63 \pm 0.974$ & $4.927 \pm 1.202$ & $0.045 \pm 0.011$ & 5350 & III,s  & sim &  PDO04 \\
V1794\,Cyg & 3.30 & $110.69 \pm 1.891$ & $4.526 \pm 3.495$ & $0.041 \pm 0.032$ & 5350 & III,s  & sim & PDO04 \\
V1794\,Cyg & 3.30 & $110.81 \pm 1.031$ & $3.782 \pm 1.604$ & $0.034 \pm 0.011$ & 5350 &  III,s  & sim & PDO04 \\
UZ\,Lib & 4.77 & $75.50 $ & $-2.039 \pm 0.227 $ & $-0.027 \pm 0.003$ & 4800 & IV,b & ccf & VKS07 \\
HU\,Vir & 10.39 & $34.38 \pm 0.15$ & $-1.01 \pm 0.23$ & $-0.029 \pm 0.005$ & 4700 & IV,b & ccf & HSK16 \\
IL\,Hya & 12.73 & $28.02 $ & $0.84 \pm 0.34$ & $0.030 \pm 0.012$ & 4500 & IV,b & sim & KW04 \\
IL\,Hya & 12.73 & $28.44 $ & $0.76 \pm 0.28$ & $0.027 \pm 0.010 $ & 4500 & IV,b & ccf & KW04 \\
IL\,Hya & 12.73 & $28.28 \pm 0.03$ & $1.43 \pm 0.15$ & $0.050 \pm 0.010$ & 4500 & IV,b & ccf & KKO14 \\
DP\,CVn & 14.01 & $25.336 \pm 0.044$ & $-0.875 \pm 0.411$ & $-0.035 \pm 0.016$ & 4600 & III,s & ccf & KKS13 \\
$\zeta$\,And & 17.77 & $19.02 \pm 0.16$ & $0.95 \pm 0.07 $ & $0.050 \pm 0.004$ & 4600 & III,b & ccf & KBS07 \\
$\zeta$\,And & 17.77 & $20.689 \pm 0.055$ & $1.138 \pm 0.045$ & $0.055 \pm 0.0022$ & 4600 & III,b & ccf & KKK12 \\
DI\,Psc & 18.07 & $19.7 \pm 0.15$ & $-1.63 \pm 0.41$ & $-0.083 \pm 0.021$ & 4600 & III,s & ccf & KKV14 \\
$\sigma$\,Gem & 19.60 & $18.26 \pm 0.07$ & $-0.840 \pm 0.13$ & $-0.046 \pm 0.01$ & 4600 & III,b & ccf & KKK15 \\
V2075\,Cyg & 22.09 & $19.297 \pm 0.074$ & $-0.772 \pm 0.386$ & $-0.040 \pm 0.02$ & 4700 & III,b & sim & WSW05 \\
V2075\,Cyg & 22.62 & $16.043 \pm 0.229$ & $0.241 \pm 0.048$ & $0.015 \pm 0.003$ & 4600 & III,b & ccf & \"OCK16 \\
KU\,Peg & 23.90 & $15.5138 \pm 0.0735$ & $0.6215 \pm 0.092$ & $0.040 \pm 0.006$ & 4440 & III,s & ccf & KKS16 \\
HK\,Lac & 24.20 & $14.876 \pm 0.062$ & $-0.744 \pm 0.744$ & $-0.05 \pm 0.05$ & 4700 & III,b & sim & WSW05  \\
IM\,Peg & 24.25 & $14.845 \pm 0.15$ & $-0.742 \pm 0.742$ & $-0.05 \pm 0.05$ & 4500 & III,b & sim & WSW05 \\
IM\,Peg & 24.65 & $14.931 \pm 0.011$ & $0.814 \pm 0.040$ & $0.054 \pm 0.003$ & 4500 & III,b & sim & MBD07 \\
V1192\,Ori & 25.30 & $13.87 \pm 0.22$ & $-1.73 \pm 0.67$ & $-0.125 \pm 0.050$ & 4500 & III,s & ccf & SKW03 \\
V1192\,Ori & 28.30 & $12.695 \pm 0.034$ & $-1.414 \pm 0.148$ & $-0.11 \pm 0.02$ & 4305 & III,s & ccf & KSO17 \\
Sun  &  25.38  &  $14.37 $ & $ 2.86 $ & $0.20$ & 5780 &  V,s & str & BVW86 \\
61\,Cyg A & 34.20 & $10.31 \pm 0.286$ & $2.292 \pm 0.573$ & $0.22 \pm 0.056$ & 4545 & V,s & sim & BJM16 \\
\hline
\end{tabular}
\flushleft
$^a$ V: dwarf, IV: subgiant, III: giant, s: single or effectively single, b: member of a close binary system\\
$^b$ sim: sheared image method, ccf: cross-correlation method, str: spot tracking technique\\
$^c$  B05: {\cite{2005MNRAS.364..137B}}, BCL05: \cite{2005MNRAS.356.1501B}, BJC04: \cite{2004MNRAS.352..589B}, DC97: {\cite{1997MNRAS.291....1D}}, MWC05: \cite{2005MNRAS.359..711M}, DCP03: \cite{2003MNRAS.345.1187D}, PDV04 \cite{2004MNRAS.348.1175P}, KSG04: {\cite{2004A&A...417.1047K}},
KWF09: \cite{2009AIPC.1094..676K},   KBH00: {\cite{2000A&A...360.1067K}}, PDO04: \cite{2004MNRAS.351..826P}, VKS07: \cite{2007AN....328.1078V}, HSK16: {\cite{2016A&A...592A.117H}}, KW04: \cite{2004PADEU..14..221K}, KKO14: \cite{2014IAUS..302..379K},  KKS13: {\cite{2013A&A...551A...2K}}, KBS07: {\cite{2007A&A...463.1071K}}, KKK12: {\cite{2012A&A...539A..50K}}, KKV14: {\cite{2014A&A...571A..74K}},  KKK15: {\cite{2015A&A...573A..98K}}, WSW05: \cite{2005AN....326..287W}, \"OCK16: {\cite{2016A&A...593A.123O}}, KKS16: {\cite{2016A&A...596A..53K}}, MBD07: \cite{2007AN....328.1047M}, SKW03: {\cite{2003A&A...408.1103S}},  KSO17: \cite{2017arXiv170801577K}, BVW86: {\cite{1986A&A...155...87B}}, BJM16: {\cite{2016A&A...594A..29B}}
\label{drstars}
\end{table*}

\section{Results}

\begin{figure*}[th]
\includegraphics[width=1.8\columnwidth]{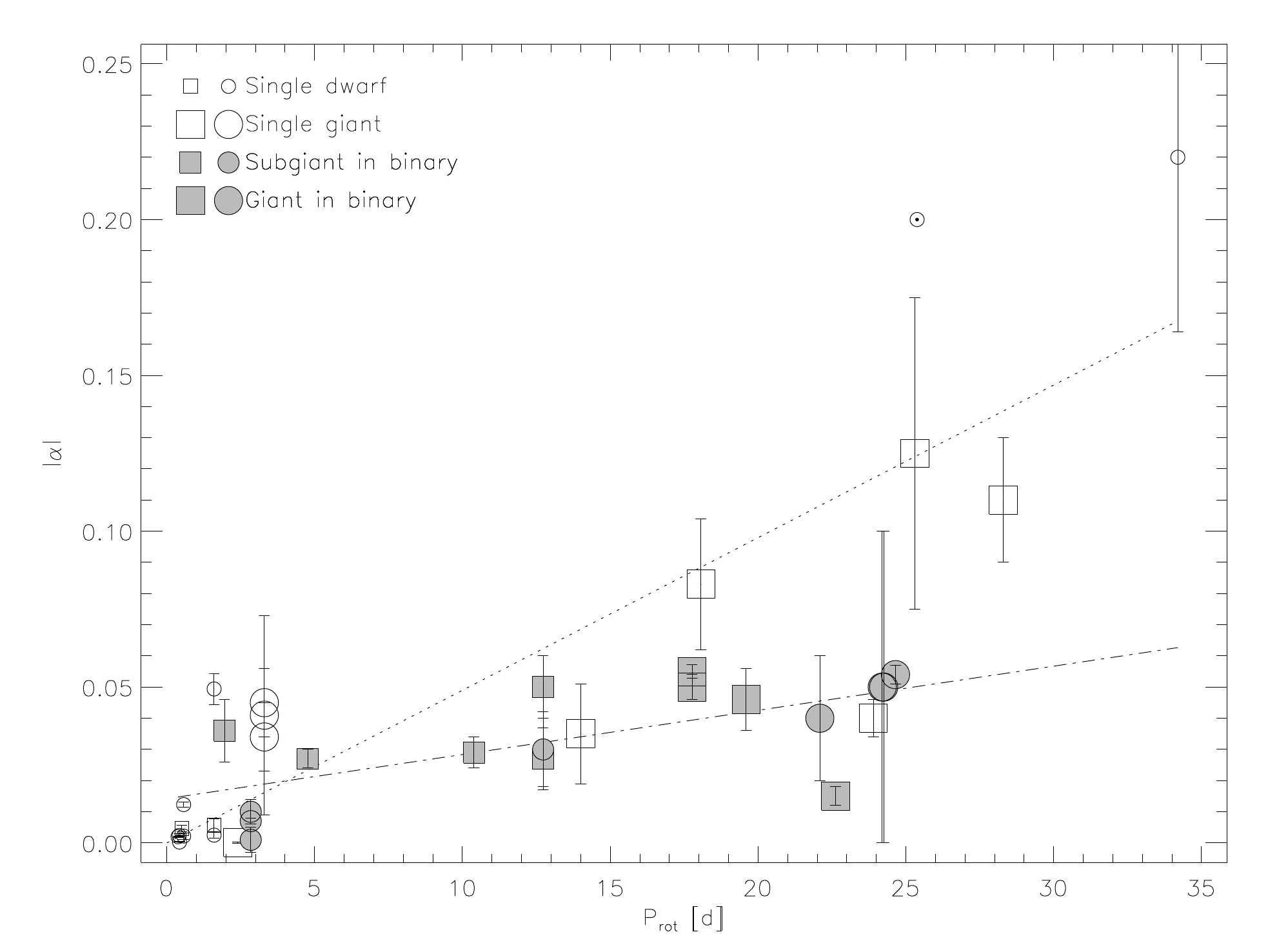}
\caption{The absolute values of the dimensionless surface shear parameter $\alpha$ from Table~\ref{Tab1} vs. rotation period. Open symbols are single stars, grey-filled symbols are members of close binary systems. Circles represent results by applying the sheared image method while rectangles are results obtained from cross-correlation technique. Dotted line represents a linear fit to the single stars while dash-dotted line fits the binary members of our sample.}
\label{fig1}
\end{figure*}

In Fig.~\ref{fig1} we plot the $\alpha=\Delta\Omega/\Omega_{\rm eq}$ dimensionless surface shear coefficient as a function of $P_{\rm rot}$. We use absolute values of $\alpha$ because in this context there is no difference of substance between solar-type and antisolar differential rotation, since the field amplification due to the $\Omega$-effect is related directly to the shear, not to its direction. The overall distribution of the datapoints reflects a trend that the longer the rotation period, the higher the surface shear coefficient \citep[cf.][]{2014SSRv..186..457K}. On the other hand, there is a significant difference between the distribution of the open symbols representing single stars and the grey-filled symbols representing stars in close binary systems. The two types of symbols are fitted (unweighted) by two different linear functions. The basic difference between the fits suggests, that the surface shear coefficient, i.e., the differential rotation is somehow confined in binary systems. The shear coefficient for close binary members grows only slowly towards the longer periods, and does not reach as high values as found for slowly rotating single stars. The fit for single stars (dotted line in  Fig.~\ref{fig1}) yields  $|\alpha|$\,$\propto$\,(0.0049$\pm$0.0001)$P_{\rm rot}$[d] while the fit for the binary members (dash-dotted line) has a restrained slope of 0.0014$\pm$0.0003. Finally we note, that plotting $|\alpha|$ as a function of $P_{\rm rot}$ is essential, since a plot similar to \citet[][Fig.~3]{2005MNRAS.357L...1B}, where $\log_{10}\Delta\Omega$ was plotted as a function of $\log_{10}\Omega_{\rm eq}$, would not reveal such striking difference between single stars and binary members. In the next section we further discuss on this  exciting result.

\begin{figure*}[th]
\includegraphics[width=1.8\columnwidth]{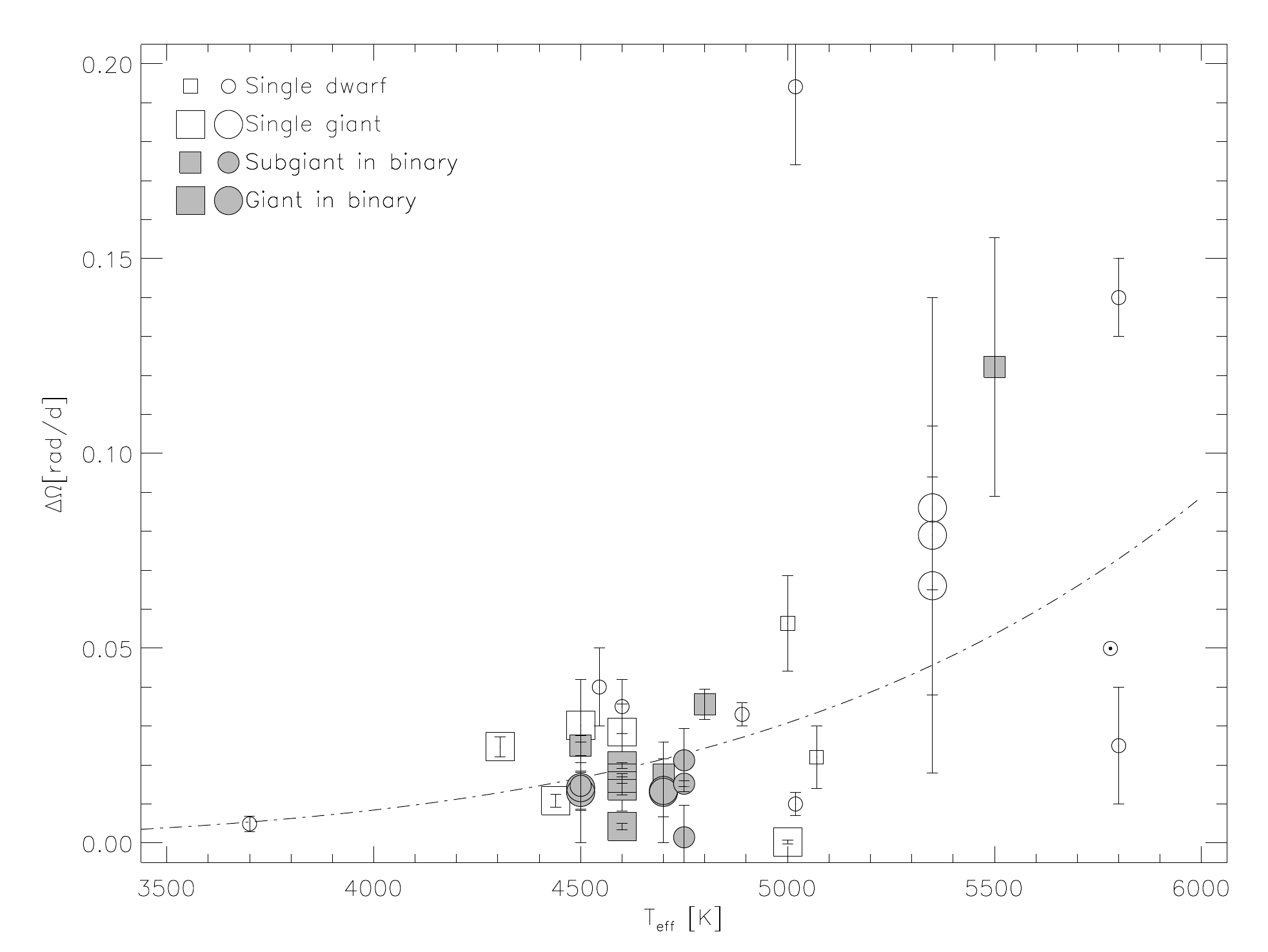}
\caption{Absolute surface shear vs. effective temperature. Dash-dotted line represents an unweighted power-law fit to all datapoints. Otherwise as in Fig.~\ref{fig1}.}
\label{fig2}
\end{figure*}

In  Fig.~\ref{fig2} we plotted $\Delta\Omega$ absolute surface shear  as a function of the effective temperature. The best fit for the unweighted data points is in agreement with the early result by \citet[][their Fig.~2]{2005MNRAS.357L...1B}, suggesting an overall trend that $\Delta\Omega$ is growing with the temperature. Assuming a power-law in the form of $\Delta\Omega\propto T_{\rm eff}^{p}$, we get $p=5.8\pm1.0$ for our sample. In the end we mention, that single stars and binaries, i.e., open and filled symbols, respectively, are not particularly separated in Fig.~\ref{fig2}.

\section{Discussion}\label{disc}

\subsection{Differential rotation of single G--K--M stars}\label{gkm}

Single late-type stars seem to follow a simple trend that fast-rotators have small relative shear, while slower rotators generally perform larger relative surface shear (with a large scatter, though).  Considering the true error bars which are very likely larger than the estimated values in Table~\ref{Tab1} (cf. Sect.~\ref{sample}), the dotted line nicely fits the observations, except one measurement of the fast-rotating LQ\,Hya at $P_{\rm rot}=1.6$\,d. There are three detections listed in Table~\ref{Tab1} for this target, two independent results among them are in agreement, claiming a weak differential rotation with $\alpha$\,$=$\,$0.0025$ and 0.0056 \citep[][respectively]{2003MNRAS.345.1187D,2004A&A...417.1047K}, while the third result, $\alpha$\,$=$\,$0.0494$ \citep[also from][]{2003MNRAS.345.1187D} is higher by an order of a magnitude compared to the other two. This fluctuation, real or not, can be associated with the changes in the kinetic energy, which was estimated to be 10 per cent of the total stellar luminosity of LQ\,Hya \citep{2003MNRAS.345.1187D}. However, this explanation is not applicable on a time scale of a year or so. It is more likely that, similarly to the case of IM\,Peg \citep{2007AN....328.1047M} already mentioned in Sect.~\ref{sample}, the fluctuating detections by \citet{2003MNRAS.345.1187D} reflect uncovered inner errors of the sheared image method applied \citep[for more discussion see Sect.~3.3.1 last paragraph in][]{2014SSRv..186..457K}. Therefore, we consider this higher shear as an erroneous detection and the smaller values as more realistic.
%{\bf
However, despite the discrepancy, we did not to exclude this datapoint from the fits  shown in  Fig.~\ref{fig1} and Fig.~\ref{fig2}.
%}

Smaller relative shear is preferred also by theory in fast-rotating dwarf stars. A strong shear would be reduced by the so-called $\Omega$-quenching, because an intensified $\Omega$-effect would yield more and more dominant toroidal fields, and the related magnetic energy would increase until the counteraction of the Lorentz-force on the plasma flow. Consequently, the differential rotation would reduce dramatically \citep[see, e.g.,][and their references]{2015SSRv..196..303B}. According to \citet{2004A&A...413.1143F}, a similar mechanism prevents the differential rotation from penetrating down into the solar radiative interior.
Note, that the differential rotation is highly quenched in fully convective M-stars as well, until the rotation is fast enough \citep{2016ApJ...833L..28Y}.
The model by \citet{1999A&A...344..911K} suggests that $\Delta\Omega/\Omega$\,$\sim$\,$\Omega^{-n}$, where $n$$\approx$$1$ for G--K dwarfs, i.e., $\Delta\Omega/\Omega$\,$\propto$\,$P_{\rm rot}$. The direct relationship between the dimensionless surface shear parameter and the equatorial period gives
$\alpha$\,$\approx$\,$P_{\rm eq}[{\rm d}]/100$ \citep[see also][]{2011A&A...530A..48K}, which agrees with the preferred smaller shear values of our fast-rotating single dwarfs. The (unweighted) linear fit  for the collected sample of single and effectively single stars in Fig.~\ref{fig1} suggests
 $\alpha$\,$\approx$\,$P_{\rm eq}[{\rm d}]/200$, 
 i.e., a two times larger denominator, still the same order. Finally, this dependence is supported also by \emph{Kepler} photometry \citep{2016MNRAS.461..497B} who found $\Delta\Omega/\Omega$\,$\propto$\,$\Omega^{-0.8}$ for G-stars and $\Delta\Omega/\Omega$\,$\propto$\,$\Omega^{-1.1}$ for K-stars, separately, i.e., with exponents close to $-1$, as suggested thereinbefore.

\subsection{Antisolar differential rotation of Li-rich single giants}

In our sample there are three single K-giants, namely DP\,CVn,  DI\,Psc, and V1192\,Ori, all three located at the RGB, rotate rapidly, and show enhanced surface Li abundance \citep{2013A&A...551A...2K,2014A&A...571A..74K,2003A&A...408.1103S}.
Moreover, these are the only known single giants that perform antisolar type surface differential rotation (cf. Table~\ref{Tab1}). The enriched surface lithium at the RGB together with antisolar differential rotation may imply a common origin of these properties, as suggested first by \citet{2014A&A...571A..74K}, but see also \citet{2017arXiv170801577K}.
The relatively rapid rotation of these objects can be explained by the interaction between the deepening convective envelope and the
fast rotating core \citep[see, e.g.][]{1989ApJ...346..303S,2010A&A...522A..10C}. 
The first dredge-up episode on the RGB is responsible for the dilution of the surface lithium, however these three giants show enriched surface lithium. This controversy may be resolved by the so-called cool-bottom processes \citep{1992ApJ...392L..71S,1995ApJ...447L..37W} where cool material from the bottom of the convective envelope is brought down to hotter layers and exposed to partial H burning to produce lithium and other light elements. Then, the lithium-rich material is transported back to the convective envelope by some deep circulation, wherefrom the lithium can reach the surface by convective mixing and/or meridional circulation.

On the other hand,
strong mixing by turbulent convection can equilibrate angular momentum, inducing antisolar differential rotation \citep{2007Icar..190..110A}.
Therefore we believe that, in case of DP\,CVn, DI\,Psc, and V1192\,Ori, antisolar differential rotation and surface lithium enrichment are indeed related attributes. In a possible scenario deep mixing is supposed to bring up fresh lithium along with angular momentum from the hotter layers into the convective envelope, and from there meridional circulation and/or turbulent convection may transport lithium as well as angular momentum towards the surface.
However, confirming this scenario needs further investigation of antisolar type differential rotation among single Li-rich giants at the RGB.

\subsection{The gravitational effect of a close companion in close binary systems}

Tidal coupling in RS\,CVn systems is responsible for maintaining fast rotation, and so makes magnetic dynamo work in the late-type companions.
Moreover, tidal effects may organize preferred longitudes, as well as latitudes of activity \citep{2007IAUS..240..442O}.
Aspherical distortion of the active component in an RS\,CVn-type binary caused by a close companion can explain the emergence of magnetic flux at preferred longitudes locked to the orbital frame, therefore the degree of the deformation could also account for disparate rotation laws \citep{2012IAUS..282..197K}. Theoretical calculations by \citet[][]{1982ApJ...253..298S} demonstrated how physical parameters of such a close binary determine the developing corotation latitude, where rotation is locked to the orbital revolution.
Tidal forces may play an important role also in inducing meridional circulation, therefore antisolar differential rotation \citep{2004AN....325..496K}.
In any way soever, tidal forces do interact with the differentially rotating convective bulk and produce a distinct difference between the relative surface shear measured on binary components compared to single stars of similar type. Indeed, according to Fig.~\ref{fig1} tidal forces seem to confine differential rotation, i.e., $|\alpha|$ may slightly be enhanced for fast-rotating binaries ($P_{\rm rot}$\,$\lesssim$\,2--3\,d) compared to single stars, but the relative shear is firmly suppressed for slower rotating ($P_{\rm rot}$\,$\gtrsim$\,10\,d) binary members.

\subsection{The effective temperature--absolute shear dependency}

Theoretical predictions agree that the surface shear is strongly dependent on spectral type \citep[e.g.,][]{1999A&A...344..911K,2005AN....326..265K,2011AN....332..933K}.  
According to our observational result (see Fig.~\ref{fig2}) the $\Delta\Omega$ absolute shear slightly increases between 3500--6000\,K towards earlier spectral types. 
(The only strikingly loose datapoint at the top of the figure is a possible misdetermination for LQ\,Hya, cf. Sect.~\ref{gkm}.)
Our fitted power-law of $\Delta\Omega$\,$\propto$\,$T_{\rm eff}^p$ with $p$\,$=$\,$5.8\pm1.0$ suggests still a strong dependency on the temperature, but with a significantly lower exponent compared to $p$\,$=$\,${8.92\pm0.31}$ from the early result of \citet[][their Fig.~2]{2005MNRAS.357L...1B}. Our result, however, is based on a three times richer and more homogeneous sample of late-type stars. Nevertheless, the lower exponent fits better the theoretical results, which is supported also by \citet[][see their Fig.~12]{2015A&A...583A..65R}, where a similar dependency was obtained for G--K spectral types using an extended \emph{Kepler} dataset.

\section{Conclusions}\label{conc}

\begin{itemize}

\item[--]We suggest alternative rotation--differential rotation relationships for late-type single stars and binaries. We find an overall trend that the differential rotation represented by the relative surface shear increases with the rotation period, however the dependency is much stronger for single stars. On the other hand, in close binary systems tidal forces confine differential rotation, and therefore, the relative shear is firmly suppressed towards longer periods. 
%{\bf
We note however, that despite the growing number of individual differential rotation measurements,
the available sample is still too small to be
representative, and more observations would definitely be essential to improve the statistics.
%}

\item[--]So far we know only three single RGB giants DP\,CVn, DI\,Psc, and V1192\,Ori that perform antisolar surface differential rotation, in addition, all three show surface lithium enrichment. We believe that, in case of these three stars, the peculiar rotation pattern and the lithium enrichment are related attributes, implying a common origin.

\item[--]We confirm that the absolute surface shear $\Delta\Omega$ is strongly dependent on the effective temperature. Our unweighted best fit suggests a power-law dependency of $\Delta\Omega$\,$\propto$\,$T_{\rm eff}^{5.8\pm1.0}$.

\end{itemize}

\acknowledgements
We thank the anonymous referee for the helpful comments that improved the paper. Authors are grateful to the Hungarian National Research, Development and Innovation Office grants OTKA K-109276 and OTKA K-113117. KV is supported by the Bolyai J\'anos Research Scholarship of the Hungarian Academy of Sciences. This paper is partly based on data obtained with the STELLA robotic telescopes in Tenerife, an AIP facility jointly operated by AIP and IAC (https://stella.aip.de/) and by the Amadeus APT jointly operated by AIP and Fairborn Observatory in Arizona. We are grateful to the ministry for research and culture of the State of Brandenburg (MWFK) and the German federal ministry for education and research (BMBF) for their continuous support .  
The authors acknowledge the support of the German \emph{Deut\-sche For\-schungs\-ge\-mein\-schaft, DFG\/} through projects KO2320/1 and STR645/1.

% Use this code if you wish to generate your bibliography with BibTeX;
% please replace first the string "an-demo" below with the name(s) of
% the BibTeX data base(s) you want to use.
% The resulting bibliography-output (the contents of the .bbl file)
% must be pasted into this file before submission.
% 
\bibliographystyle{an}
\bibliography{kovarietal}
% 
% Replace the following example bibliography with your references
% before submission:
%\begin{thebibliography}{}
%  \bibitem{} Author1, A.B. \& Author2, C.D. 2001, \an, 322, 1
%  \bibitem{} Author3, E.F., Author4, G.H., \& Author5, I. 2001, \aap, 322, 10
%  \bibitem{} Author5, I. 2001, \mnras, 322, 20
%  \bibitem{} Author6, J. 2001, \apj, 322, 30
%\end{thebibliography}

\end{document}